# Evaluating the impact of word embeddings on similarity scoring in practical information retrieval

Niall McCarroll[1], Kevin Curran[1], Eugene McNamee[2], Angela Clist[3], Andrew Brammer[3]

[1] Ulster University, School of Computing, Engineering and Intelligent Systems
Derry, Northern Ireland
n.mccarroll@ulster.ac.uk

[2] Ulster University, School of Law,
Jordanstown, Newtownabbey, Northern Ireland

[3] A&O Sherman
Donegall Quay, Belfast, Northern Ireland

## Abstract

*Search behaviour is characterised using synonymy and polysemy as users often want to search information based on meaning. Semantic representation strategies represent a move towards richer associative connections that can adequately capture this complex usage of language. Vector Space Modelling (VSM) and neural word embeddings play a crucial role in modern machine learning and Natural Language Processing (NLP) pipelines. Embeddings use distributional semantics to represent words, sentences, paragraphs or entire documents as vectors in high dimensional spaces. This can be leveraged by Information Retrieval (IR) systems to exploit the semantic relatedness between queries and answers.*

*This paper evaluates an alternative approach to measuring query-statement similarity that moves away from the common similarity measure of centroids of neural word embeddings. Motivated by the Word Mover's Distance (WMD) model, similarity is evaluated using the distance between individual words of queries and statements. Results from ranked query and response statements demonstrate significant gains in accuracy using the combined approach of similarity ranking through WMD with the word embedding techniques. The top performing WMD + GloVe combination outperforms all other state-of-the-art retrieval models including Doc2Vec and the baseline LSA model. Along with the significant gains in performance of similarity ranking through WMD, we conclude that the use of pre-trained word embeddings, trained on vast amounts of data, result in domain agnostic language processing solutions that are portable to diverse business use-cases.*

## 1. Introduction

Textual artefacts are now being generated faster than at any stage of in the development of information technology (Grainger, T.; Potter 2014). One of the main challenges for any large business is how to manage and navigate through these rapidly growing repositories of text-centric data (O'Connell et al., 2025a; O'Connell et al., 2025b). To address this adequately, Information Retrieval (IR) strategies have been developed for the searching, organisation and representation of unstructured knowledge and this has become central to the entire document management process (Manning et al. 2009). Information retrieval can be considered from various levels of processing including single words or sentences to paragraphs and full document retrieval. The most important challenge is the retrieval of documents that are relevant to user queries. The ability to rank results by query relevance is a key aspect of Information Retrieval and it is this function that differentiates IR from other types of database queries which are sorted by one or more table columns. IR systems such as search engines return results sorted into descending order based on a score that designates the strength of match between the query and the returned document. If we consider business scenarios where potentially millions of documents are to be searched and processed, ranked retrieval has important implications for efficiency as it prevents the user from being overloaded with results that are impossible to navigate and consume (Růžička et al. 2017). There are several ways of calculating and influencing relevancy ranking. For certain documents, fields or terms can be 'boosted' by assigning more weight to them or the recency of a document can elevate it higher in the rankings compared to older artefacts. The usage clicks or views that a document receives can also be factored into its weighting (McCarroll et al., 2025).

However, the focus of this investigation is improving information discovery retrieval through measures of semantic relevancy. Semantic relevancy is fundamental to information retrieval. Words can belong to multiple concepts, with varying strengths of association with each concept. Semantic or conceptual learning in language processing is a complex concept that goes beyond traditional lexical matching techniques, where judgements of likeness are measured in terms of factors such as synonymy, polysemy, redundancy, entailment and paraphrasing (Islam & Inkpen 2008). The concept of synonyms or semantic heterogeneity in language is one in which the same real-world entity can be represented using different linguistic terms. For example, the concept of word 'beautiful' can also be conveyed with the words 'attractive', 'pretty', 'lovely' and 'stunning'. The ability to manage synonymy has important implications for querying data from multiple sources, as well as the cleansing and mining of data (Cohen 2000; Schallehn et al. 2004; Madhavan et al. 2005). Polysemy is the capacity of a word or phrase to have multiple meanings. For example, the word 'solution' has two different meaning when we compare its use in the sentence – 'work out the solution in your head' to its use in the sentence – 'heat the solution to 75° Celsius' (Salton & McGill 1983; Blair & Maron 1985; Dumais et al. 1988).

Many traditional keyword-based information retrieval systems rely on statistical term overlap. This direct mapping of a query with indexed terms or statements, suffers from lexical gaps when we consider that conceptually identical or similar statements can be expressed in a variety of ways and, conversely, completely unrelated concepts can be textually quite similar. As a result, these traditional information retrieval systems often miss relevant documents or return irrelevant ones that contain different terms than the query, even with the use of query expansion (Brokos et al. 2016). In addition to this, ambiguity and the complexity of expression contribute significantly to this problem (Dabney 1986). Users cannot predict the terms and phrases that will be used in the documents they will ultimately find useful and relevant (Blair & Maron 1985). The ability to identify related terms outside of the keyword range is also important for handling partial queries or scenarios where the user is conducting exploratory searches based on minimal or imprecise details (van Opijnen & Santos 2017). These complexities have driven the development of Natural Language Processing (NLP) over pure text-based search solutions, in an effort to interpret language in a more complex and meaningful way (Maxwell & Schafer 2008; Mikolov, Le, et al. 2013). Semantic analysis goes deeper than word to object associations to uncover the links between the larger pool of words that can be attributed to each object. There are a number of Machine Learning and statistical strategies that can be employed to estimate and uncover this underlying latent structure of meaning. These algorithms work to organise text into a semantic structure that can be leveraged to maximise the representation and retrieval of information thus facilitating information navigation (Dumais et al. 1988). Semantic similarity matching has been shown to improve recall and precision. and has a wide range of applications within Natural Language Processing including text summarisation (Erkan & Radev 2004; Lin & Hovy 2003), evaluation of text coherence (Wegrzyn-Wolska & Szczepaniak 2005; Lapata & Barzilay 2005), word sense disambiguation (Lesk 1986; Schütze 1998), text categorisation and relevance feedback (Liu & Guo 2005; Ko et al. 2004). Semantic learning is also considered to be one of the best techniques for improving the effectiveness of information retrieval [4] [24] (Grainger, T.; Potter 2014). Word and paragraph embedding algorithms such as Word2Vec (Mikolov, Chen, et al. 2013), Doc2Vec (Le & Mikolov 2014), GloVe (Pennington et al. 2014) and FastText (Bojanowski et al. 2016) have emerged as leading approaches to modelling the semantic relations between terms in various Natural Language Processing pipelines. This paper evaluates existing and newly proposed models that integrate these pre-trained neural word vector embeddings into the matching and ranking phases of the information retrieval pipeline.

This research makes the following important contributions. Firstly, this work represents one of the first investigations to carry out comparative evaluation of techniques that combine similarity ranking through Word Mover's Distance (WMD) with each of the Word2Vec, FastText and GloVe word embedding algorithms. WMD measures query-statement similarity based on an evaluation of the distance between individual words, as opposed to the common similarity measure that uses query-statement centroids of word embeddings. Secondly, the retrieval models' robustness to statement length is investigated by using multi-length query-statement matches ranging from single short sentences to multi-sentence paragraphs. Results from the comparative analysis clearly show that implementing a two-pronged approach that combines GloVe word embeddings with

WMD similarity ranking achieves robust retrieval and ranking accuracy. This robust accuracy was achieved using an unsupervised, pre-trained word vector model that does not require domain specific ontologies or manually labelled training data. Consequently, it was concluded that these techniques do not necessitate domain specific training and can be easily ported to any number of different business use cases such as legal, financial and medical applications.

The remainder of this paper is organised as follows: Section Two discusses related work and reviews existing embedding-based approaches for conceptual search strategies. Section Three details the methodology and experimental set-up that was implemented to evaluate the different word embedding models on a practical information retrieval task. Section Four discusses results and analysis of the query-statement matching trials before concluding with a detailed discussion of findings and future work in Section Five.

## 2. Semantic Representation and Information Retrieval

The challenge of matching documents or statements based on their textual descriptions remains an active area of Information Retrieval research. Moving away from the traditional approach of counting query term occurrences in the target search text, many latent semantic and more recent neural embedding methods have been proposed to bridge the gap caused by linguistic and vocabulary-related mismatches and differences. Algorithmic relevance is at the computational core of Information Retrieval and concerns the relationship between information objects and user queries based on some measure of similarity between them. The 'gold standard' of algorithmic relevance performance is that the search engine query should retrieve specified sets of information objects, measured as recall with a minimum number of false positives, measured as precision (Grainger, T.; Potter 2014). Previous efforts to address the diversity of human language in Information Retrieval have included: the augmentation of user's original queries with intermediary search keys of related terms; restricting allowable vocabulary; or constructing domain specific ontologies to reflect language that is common to a specific task. These methods proved to be both labour intensive and inadequate (Furnas et al. 1983; Furnas et al. 1987; Dumais et al. 1988).

Traditional vector-based algorithms such as Okapi BM25 (Robertson et al. 1992) and TF-IDF (Term Frequency – Inverse Document Frequency) (Salton & Buckley 1988) count term occurrences and utilise bag-of-words representations reweighted by inverse document frequency. In the bag-of-words approach, text documents are represented by isolated keyword terms that have no syntactic or semantic context or relation to other terms in the model. While these approaches are strong baselines, they fail to adequately represent complex text objects such as sentences and paragraphs because all relationships and term dependencies are lost. In particular they struggle with vocabulary mismatch, where semantically related terms are disregarded (Manning et al. 2009). Recall for these term-matching systems is generally below acceptable levels because they depend on exact word or phrase matching. Therefore, they are only useful for full text matching activity such as keyword or short phrase retrieval. Issues arise when the requester uses different words from the source information leading to imprecise and incomplete matching with unrelated and irrelevant text responses being returned, and query relevant text objects being missed (Mikolov, Chen, et al. 2013; Mohan et al. 2011).

The use of latent semantic structures to facilitate Information Retrieval is well established. Semantic analysis models such as Latent Semantic Analysis (Deerwester et al. 1990) and the probabilistic topic model, Latent Dirichlet Allocation (LDA) (Blei et al. 2003) map dense vector representations onto low-dimensional subspace. Latent Semantic Analysis (LSA) or Latent Semantic Indexing (LSI) (Landauer & Dumais 1997; Landauer et al. 1998) is a distributional semantics technique and is one of the best known methods for corpus-based similarity comparisons. It works by analysing a large corpus of natural language to derive high-dimensional linear associations that represent the similarity of words and other text objects. LSA quantifies and groups semantic similarities between text objects using the assumption that words which are close in meaning will occur in similar text segments or linguistic contexts. This is achieved by representing the documents to be queried as a huge term-document matrix containing word counts per paragraph (rows represent unique words and columns represent each paragraph). Singular Value Decomposition (SVD) is used to perform dimensionality reduction on the large number of TF-IDF feature vectors by reducing the number of rows while preserving the similarity structure among columns. As a result, each sentence is represented as a vector in reduced-dimensional space and the resulting features are much more informative. This is how LSA increases the classification accuracy over the plain TF-IDF approach. Similarity can then be measured by the cosine of the angle between their corresponding row vectors. Sentences or text objects such as words that refer to the same concept will be mapped closer together in the vector space leading to higher cosine values (closer to 1) (Landauer et al. 1998).

Since LSA calculates similarity based on context, it does not require queries and target statements or documents to contain common words and, therefore, outperforms traditional vector space models in measures of synonymy. It is easy to implement and efficient at run-time as it only involves decomposing the term-document matrix. However, as a global bag-of-words approach, it fails to efficiently leverage lower level syntactic and statistical information that exposes the links between component vectors. As a consequence, LSA tends to be more appropriate for similarity matching of longer texts as opposed to key word matching (Islam & Inkpen 2008; Dumais et al. 1988). Therefore, even though synonymy is addressed to a certain extent, methods such as LSA and LDA exhibit inferior performance to classic ranking algorithms such as BM25 in various standardised IR tasks. It should, however, be noted that these standardised IR tests were not developed to measure semantic performance (Pennington et al. 2014; Kim et al. 2017).

A further criticism of these earlier models is that they tend to develop domain dependency to the datasets that they are trained on, making it more difficult to adapt to other applications or knowledge domains. This is in contrast to natural human language observations where sentence meaning easily adapts between domains (Dumais et al. 1988; Kim et al. 2017).

## 2.1 Neural Word Embeddings

To bridge the lexical gap caused by linguistic difference and to adequately respond to the challenge of representing documents semantically has prompted the development of advanced representation techniques such as Distributional Semantic Models (DSMs) that signify a move away from simple syntactic matching mechanisms, to a more complex combinations of syntactic and semantic parsing to enhance search recall and expressiveness (Uren et al. 2007). Computational linguistics has reliably shown that contextual information affords a good approximation of word meaning, as semantically related words will tend to occur in similar contextual distributions (Miller & Charles 1991; Baroni et al. 2014). DSMs employ vectors to track these contexts in which target terms appear and store them as meaning representations. Geometric techniques are then applied to these vectors to measure the similarity in meaning between search and target phrases (Baroni et al. 2014). From the statistical neural net language models introduced by Bengio et al. (Bengio et al. 2003) evolved word embeddings learned by neural networks. Like LSA and LDA, these neural word embeddings are distributed, low-dimensional word vector representations, which are learned from the raw text data sources. The word embeddings provide latent semantic representations for the heterogeneous textual data from which they are generated (Mikolov, Chen, et al. 2013). In contrast to the LSA or LDA representations that utilise co-occurrences of words, word embeddings learn semantic word vectors to predict context words, thus capturing both the syntactic and semantic relations between the collections of words that constitute a sentence or paragraph (Pennington et al. 2014; Schnabel et al. 2015).

Both neural word embedding and co-occurrence approaches have been applied to lexical semantic tasks such as synonym detection, topic clustering, concept categorisation and semantic relatedness, but the co-occurrence algorithms are generally outperformed by the neural word embedding approaches that also require less parameter optimisation (Baroni et al. 2014; Schnabel et al. 2015). The successful application of word embeddings to ad-hoc Information Retrieval (IR) tasks remains a key area of research activity. Kusner et al. (Kusner et al. 2015) provide a useful example of the benefits of word embeddings when they describe the scenario of querying a document with a high occurrence of the term 'automobile' with a query term 'car'. Techniques such as TF-IDF will score the document relatively low since the term 'car' does not feature prevalently in the document, whereas a word embedding approach will score the document higher because the vector representation for 'automobile' and 'car' are close to each other in the embedding space. Galke et al. (2017) provide further evidence that an aggregated word vector approach outperforms the TF-IDF baseline in a range of practical information retrieval experiments. Such has been the success and impact of neural word embeddings in NLP tasks, that they are now recognised as being the main driver for the renewed interest and breakout of NLP in the past few years (Goth 2016). These neural word embeddings have become the default representations in many text processing pipelines and neural network architectures, and function as the first layer of pre-processing, converting raw word tokens into more useful representations (Bengio et al. 2003; Bengio et al. 2013; Goth 2016; Levy & Goldberg 2014). Mikolov et al. (2013) proposed Word2Vec, an unsupervised, shallow neural network based skip-gram model in which word vector representations are learned by reconstructing each word's context through an efficient training algorithm that does not utilise dense matrix manipulation (Fig 1A). Recall is greatly boosted over keyword matching techniques as Word2vec works to bring semantically synonymous vectors closer together in the embedded space. The original implementation of Word2Vec uses centroids of word vectors and cosine similarity to evaluate document relatedness.

FastText is a Facebook Artificial Intelligence Research (FAIR) open source library for natural language processing (Bojanowski et al. 2016). The FastText approach combines a number of robust Machine Learning and Natural Language Processing techniques including bag of words and bag of n-grams, as well as using sub-word information, and sharing information across classes through a hidden representation to achieve efficient text classification. FastText is on par with some deep learning frameworks in terms of accuracy in learning word vector representations but is much faster to train, making it much more scalable to larger datasets.

FastText extends the capabilities of Word2Vec as it not only learns vectors for complete words but it also learns vectors for the n-grams that each word is composed of. For example, the word vector "apple" is a sum of the vectors of the n-grams "<ap", "app", "appl", "apple", "apple>", "ppl", "pple", "pple>", "ple", "ple>", "le>". These low dimension N-gram vector representations are shared across all classifiers enabling the cross category sharing of word and sub-word information. Both the target word vectors and their constituent N-gram vectors are used in each training stage in FastText. Despite the increased computation and training time that sub-word n-gram processing requires, combining word vectors with embedded sub-word information has been shown to increase FastText's accuracy over Word2Vec when handling uncommon words or words that are out-of-vocabulary. Even if words are unusual or have not appeared in a training corpus, it is likely that their component n-grams are familiar and shared with many existing words, allowing the algorithm to synthesise consistent vectors for the word that is fit for purpose. For example the n-gram 'hav' in the word 'have' shares the same n-gram in the word 'behave' (Bojanowski et al. 2016).

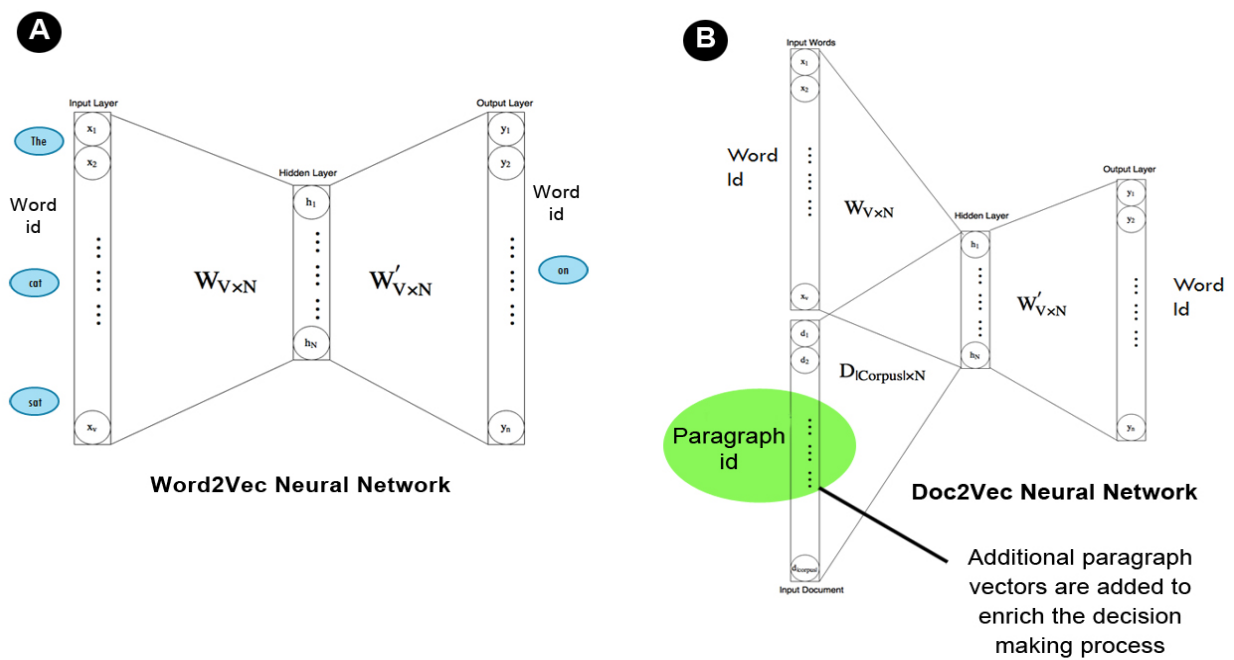

*Figure 1 – A: The Word2Vec Neural Network architecture. B: The Doc2Vec Neural Network Architecture*

Pennington et al (Pennington et al. 2014) introduced the count-based GloVe (Global Vectors) model to learn vectors through dimensionality reduction on a co-occurrence counts matrix. The large co-occurrence matrix of words (rows) and context (columns) maps how frequently each word appears in certain context in a large corpus. The large matrix is factorized to yield a lower-dimensional (word * features) matrix in which each row yields a vector representation for each word. This is achieved through reconstruction loss where lower-dimensional representations are found that account for the majority of variance in the high-dimensional data. The GloVe technique of generating word vector spaces with meaningful substructure has been found to achieve state-of-the-art performance on several text processing experiments, such as word similarity tasks, word analogy tasks and Named Entity Recognition (NER). A number of cross comparison studies evaluating the strengths and weaknesses of Word2Vec, FastText and GloVe have generally concluded that they are overall comparable to each other, although performance does vary depending on the task and the length of text both in the queries and target documents (Muneeb et al. 2015; Cao & Lu 2017). A limitation of Word2Vec, FastText and GloVe is that by only encoding vector representations for single words they do not capture the enriched data that is provided through more complex, multiple word structures such as sentences or paragraphs. Taking the concept of context beyond single words, Le and Mikolov (2014) further extended the Word2Vec framework with a semantic enriching strategy that learns fixed-length feature representations from variable-length segments of text, such as sentences, paragraphs, and entire documents. Doc2Vec (or paragraph vectors) represents a combined approach where each word is mapped to a unique vector (word embedding) as well as each paragraph being mapped to a unique vector (paragraph embedding) (See Fig 1B). This is achieved in the network through additional input nodes that represent statements or paragraphs to the network. While paragraph vectors are unique to each paragraph, the word vectors are shared across all paragraphs. For example, the vector for the word 'financial' will be the same for all paragraphs within a document. Their experiments found that this Doc2Vec framework, with its enhanced decision-making capability, proved successful in information retrieval and sentiment analysis tasks, reporting less retrieval and classification error than comparable algorithms that employed Word Centroid Distance similarity measures. If we consider real-world applications, the ability to extract meaning from larger units of text, such as multiple sentences, can be extremely advantageous when we consider the complex nature of many business documents, where adequately conveying the meaning of highly involved or technical concepts may require a greater volume of in-depth descriptive prose.

*2.2 Similarity Measures*

The semantic-based Information Retrieval process can be considered an implementation of two phases. The first stage is the processing of text to semantic vector space which has already been discussed. The second stage involves the ranking of candidates through some mechanism of similarity comparison. A well-established approach to computing similarity between sentences or documents is to evaluate the cosine similarity or inner product of the centroids of word embeddings (generated from techniques such as Word2Vec or GloVe etc.) (Nalisnick, E.; Mitra, B.; Craswell, N. Caruana 2016). These document similarity measures have been useful for general clustering and classification of overall topics at a document level. However, simple centroid approximation is regarded as insufficient for calculating the distances between queries and target statements or documents (Kusner et al. 2015). As queries tend to be short compared to the documents they are being compared against, a lossy centroid approach that calculates the average distance between a query and a document will be less accurate than an approach that searches directly for the query words (Kim et al. 2017). The centroid approach also struggles with documents that consist of multiple different topics.

Word Mover's Distance (WMD) (Kusner et al. 2015) emerged from a statistical approach known as Earth Mover's Distance (EMD) (Rubner et al. 2000) which has been successfully applied to computer vision tasks such as image comparison. EMD measures the distance between two probability distributions over a region and, similarly, the constituent word-vectors of sentences or paragraphs can be considered as distributions or 'piles of meaning' around their individual vector coordinates. WMD has been specifically developed to measure the similarity between two bodies of text (sentences / paragraphs) by calculating the minimum 'travelling distance' between text objects (sentences) as a measure of the sum-of-distances (cosine distance) or effort it takes to move from one word vector pile configuration to another. Several recent studies take advantage of the dual process of leveraging the benefits of word embeddings with WMD similarity ranking. For computing similarities between documents, the WMD approach has reported lower classification errors when used in conjunction with distance-based classifiers (Kusner et al. 2015). Combing neural word embeddings with a WMD similarity mechanism was also found to outperform a BM25 ranking system on the TREC 2006 and 2007 Genomics benchmark sets (Hersh et al. 2006; Hersh, et al. 2007) using solely semantic comparison as the ranking feature (Kim et al. 2017).

# 3. Methodology

The objective of this work is to compare several semantic representation and information retrieval strategies. For baseline comparison, a Latent Semantic Analysis (LSA) model, which represents a Global Matrix Factorization approach, is tested along with Local Context Window (skip-gram) algorithms of Doc2Vec, Word2Vec, FastText, GloVE. As an alternative approach to measuring query-statement similarity using neural word embeddings, this work is particularly motivated by the Word Movers' Distance (WMD) model (Kusner et al. 2015). Thus, moving away from a global common similarity measure using query-statement centroids of word embeddings, the proposed model evaluates similarity using the distance between individual words of queries and statements.

To the best of the authors' knowledge this is one of the first studies to use and evaluate a combination of WMD along with these different neural words embedding techniques for semantic information retrieval and similarity ranking. This section describes the experimental setup, evaluation dataset, pre-processing and evaluation metrics that were implemented to assess the accuracy of each semantic-based information retrieval system. The information retrieval performance of four state-of-the-art semantic representation techniques - Word2Vec, Doc2Vec, FastText and GloVe - is compared with a baseline performance of a traditional vector-based LSA model. The word embeddings from Word2Vec, FastText and GloVe were processed with a Word Movers' Distance (WMD) document similarity algorithm to assess the effects on statement similarity rankings compared to three Doc2Vec models that vary word order and contextual analysis.

**Dataset -** The evaluation dataset was prepared from a publically available 2013 IPO Prospectus for Foxtons Estate Agents of London (Foxtons 2013). The Prospectus consists of 223 pages of company and financial data, totalling 141,171 words over 8,127 individual statements or paragraphs. 12 Statements were taken from the Prospectus to be used in 12 separate testing trials. For each of the 12 statements a set of queries were developed that were syntactic variations of the initial statement. Each search trial query began with the original statement taken from the prospectus. The additional query statements are variations of the original statement in that textually, they are constructed differently but they convey similar meaning.

Each retrieval model's sensitivity to statement length was evaluated by including statements of varying lengths from a single sentence containing 10 words to multiple sentence statement composed of 8 sentences and 215 words for the retrieval tasks. Additionally, some search queries contained only a portion of the original statement to test if the models would return the paragraph within which the statement snippet occurs. Altogether there were 59 separate search statements across the 12 trials. Two examples of the 12 statements and their query variations can be seen below:

## *Example 1*

Target Statement: ***Michael Brown is the Chief Executive Officer of the Company.***

*Query 1:* Michael Brown is the Chief Executive Officer of the Company.

*Query 2:* The Chief Executive Officer of the Company is Michael Brown

*Query 3*: Chief Executive Officer of the Company

*Query 4:* Michael Brown

*Query 5:* Chief Executive Officer

*Query 1 is the original statement and Query 2 swaps the name of person with his job title. Query 3 searches for last section of the statement only, whereas Query 4 is restricted to the name of CEO. Finally, Query 5 restricts the search to the actual job title.*

## *Example 2*

Target Statement: ***Operating profit decreased by £0.1 million to £31.4 million between 2010 and 2012, and increased by £3.9 million to £17.5 million between the six months ended 30 June 2012 and the six months ended 30 June 2013.***

*Query 1:* Operating profit decreased by £0.1 million to £31.4 million between 2010 and 2012, and increased by £3.9 million to £17.5 million between the six months ended 30 June 2012 and the six months ended 30 June 2013.

*Query 2:* Operating profit increased by £3.9 million to £17.5 million between the six months ended 30 June 2012 and the six months ended 30 June 2013, and decreased by £0.1 million to £31.4 million between 2010 and 2012

*Query 3:* Operating profit decreased by £0.1 million between 2010 and 2012, and increased by £3.9 million between the six months ended 30 June 2012 and the six months ended 30 June 2013.

*Query 4:* Operating profit decreased to £31.4 million between 2010 and 2012, and increased to £17.5 million between the six months ended 30 June 2012 and the six months ended 30 June 2013.

*Query 5:* Operating profit decreased between 2010 and 2012, and increased between the six months ended 30 June 2012 and the six months ended 30 June 2013.

*Query 1 is the original statement and Query 2 swaps the order of the main subsections of the statement. In Query 3, only profit increase and decrease figures are included whereas in Query 4, the statement only includes the final figures. Finally, in Query 5, no financial figures are included.*

**Text Pre-processing -** To ensure fair comparison, the same pre-processing steps were observed for all information retrieval models. Firstly, the raw text string was converted to lower case. The string was subsequently tokenised by splitting it into the sub-unit words and paragraph returns were treated as delimiters to specify the text boundaries. The final stage of pre-processing was to remove all common English stop words. As a number of the models leverage the use of sub-words or sub-grams it was decided that stemming and lemmatization would not be applied to the text.

**LSA baseline model**: The document index for the LSA model was created using the IPO prospectus as the training material to generate the word-to-paragraph matrices. The similarity of the query vectors to the vectors in the document space were measured using cosine similarity.

**Embedding models** - Based on experimental recommendations from Kusner et al. (2015) and Mikilov et al. (2013), it was decided to employ robust pre-trained general purpose word embedding models as opposed to corpus or domain specific frameworks. These models are trained over vast amounts of data, thus providing a wide diversity of contexts for each word during training. This ensured that the models were not over sensitive to the test dataset.

*Word2Vec* vectors were generated from the Google News dataset (300 dimensions trained on $100 \cdot 10^9$ tokens with a vocabulary size of $3 \cdot 10^6$) (Google).

*GloVe* vectors were generated from the Common Crawl dataset (300 dimensions trained on $840 \cdot 10^9$ tokens with a vocabulary size of $2.2 \cdot 10^6$) (Common Crawl).

*FastText* vectors were generated from the Common Crawl dataset (300 dimensions trained on $840 \cdot 10^9$ tokens with a vocabulary size of $2.2 \cdot 10^6$) (Facebook).

*Doc2Vec* - There are three variations in the Doc2Vec experimental set-up:

- Paragraph Vectors – Distributed Memory Model (PV-DM)
- Paragraph Vectors – Distributed Bag of Words Model (PV-DBOW)
- Paragraph Vectors + Distributed Bag of Words Model (PV + DBOW)

Each algorithm variation processes text in a different way placing different emphasis on word order and contextual analysis. These three variations will each be evaluated for strengths and weaknesses.

*Paragraph Vectors Distributed Memory Model (PV-DM)* - This is the original Doc2Vec parameters where the additional paragraph vector acts as a distributed memory store of what is missing from the current context or the topic of the paragraph, and functions as an additional pseudo-word ranging over the entire text (sentence, paragraph or document) participating in all sliding window samples of Word Vectors. In the PV-DM model the order of words is important and many contributors believe this to be an advantage over the 'bag-of-words' approach as it preserves more information about the paragraph (Le & Mikolov 2014).

*Paragraph Vector – Distributed Bag of Words Model (PV-DBOW)* - Unlike the PV-DM model, the PV-DBOW model adopts a 'bag of words' approach where word order is irrelevant, and no Word Vectors are trained. Instead, Paragraph Vectors are trained to predict words randomly sampled from the paragraph in the output without using local neighbouring words. This simplified approach is more efficient as it requires less storage data. It is also a slightly more flexible approach than PV-DM as word order is not considered important. It is more likely that this model will recognise semantically similar but syntactically different text (Le & Mikolov 2014).

*Paragraph Vectors + Distributed Bag of Words Model (PV + DBOW)* - To produce more consistent results across multiple tasks, Le & Mikolov (Le & Mikolov 2014) recommend generating paragraph vectors that are a combination of two vectors: one learned by the standard PV-DM model and one learned by PV-DBOW approach. This combinatorial approach of the previous two methods implements simultaneous training of both Paragraph Vectors over the whole text and skip-gram Word Vectors (bag-of-words) over each sliding context window. This approach is slower as the additional training is computationally expensive, however the benefits of placing both Word Vectors and Paragraph Vectors into the same space enhances the expressiveness and interpretability of the Paragraph Vectors due to their closeness to words of known meanings (Le & Mikolov 2014).

For the Doc2Vec models, the word embeddings and paragraph centroids are calculated for the prospectus training set. The centroid is then computed for each query and the statements or paragraphs with the top 20 nearest centroids (in terms of cosine similarity) to the query are retrieved.

**WMD and similarity ranking**

As an alternative approach to measuring similarity with query-statement centroids of word embeddings, this research evaluates WMD (Kusner et al. 2015) as a means of evaluating similarity through the distance between individual words of queries and statements. It was decided to test the WMD model across several well-established pre-trained vector libraries to determine the best combination for effective semantic information retrieval. The three model combinations were as follows:

- WMD + Word2vec
- WMD + FastText
- WMD + GloVe

The text from the IPO Prospectus and the query text used to search the document was encoded as vectors through the above word embedding techniques before applying query-statement similarity comparisons with the WMD architecture. Applying the different similarity comparisons across all techniques (LSA, Doc2Vec and WMD variations), a set of relevance scores for each query-statement pair is generated. The relevancy scores are ranked and consideration is limited to the top 20 ranked statements.

## 4. Evaluation

To evaluate the robustness and accuracy of each semantic retrieval method, three separate comparisons were made regarding the ranking of correct statement matches. The first comparison assessed the number of correct matches that each model returned within the top 20 ranked positions when similarity ranking was applied. The second comparison assesses the number of correct matches that are ranked within the top three positions when similarity metrics are applied. Finally, the last evaluation assesses how many target statements each model returns to the number one, top ranked position when similarity ranking is applied.

Table 1: The number and percentage of correct statement matches returned within the top 20 similarity ranked positions by each semantic retrieval model across 12 query-statement trials totalling 59 comparison statements.

| System | # Correct Statement Matches | % Correct Statement Matches |
| --- | --- | --- |
| **WMD-GloVe** | **59** | **100%** |
| **WMD-FastText** | **59** | **100%** |
| **WMD-Word2Vec** | **59** | **100%** |
| PV + DBOW | 53 | 89.83% |
| PV-DBOW | 40 | 67.8% |
| PV-DM | 33 | 55.93% |
| LSA | 11 | 18.64% |

The WMD-GloVe, WMD-FastText and WMD-Word2Vec systems outperform the other text comparison systems returning 100% correct matches within the top 20 ranked results. The only system coming close to this performance is the Doc2Vec PV + DBOW method which returns 89.83% (53/59) correct matches. Upon further analysis it was discovered that all three systems return 100% correct statement matches within the top 10 results. Amongst the Doc2Vec models, the PV-DBOW version of Doc2Vec outperforms the original PV_DM model returning 40/59 (67.8% return rate) statement matches compared to 33/59 (55.93% return rate). However, the PV + DBOW Doc2Vec model is the clear winner with a return rate of 89.83 % returning 53 correct statement matches out of a possible 59. This stronger performance of the PV + DBOW Doc2Vec model is in accordance the findings and recommendations of Le & Mikolov (2014). The combination of both low-level skip-gram Word Vectors with higher level Paragraph Vectors harnesses greater semantic expressiveness when they work together in the same distribution space.

The superior performance of the 'bag-of-words' Doc2Vec PV-DBOW approach compared to the PV-DM model indicates that it is much more flexible to the change of word ordering that occurs in many of the trials compared to the more rigid PV-DM sliding paragraph window method. In this case, the benefit of preserving additional information about the paragraph through the sliding window technique is outweighed by the 'bag-of-word' approach where word order is not important. This failure to generalise to rephrased paragraphs indicates that PV-DM may be too restrictive when we are trying to match statements that are semantically similar yet textually different.

The LSA model achieves a very poor return of 11 correct statement matches (18.64%) in the top 20 relevancy ranked positions. Comparing this baseline model to the more advanced word and paragraph embedding approaches can be considered in broader terms as a comparison between two main approaches to learning and generating word vectors. LSA is a Global Matrix Factorization method that processes text at the higher document level compared to the Local Context Window (skip-gram) approaches such as the Doc2Vev variations and the word embedding algorithms. The poor performance of the LSA model reinforces evidence that global approaches fail to efficiently leverage the lower level statistical information that exposes the links between component vectors. The Global techniques tend to work better at document level processing such as topic clustering or classification. Indeed, from informal qualitative analysis of the top 20 results returned for each LSA query, there is no obvious semantic consistency or relatedness in the content or themes of the returned statements. All WMD systems have 100% record for returning top 20 ranked statement matches.

*Table 2: The number and percentage of correct statement matches returned within the top three similarity ranking positions by each semantic retrieval model across 12 query-statement trials totalling 59 comparison statements.*

| System | # Correct Statement Matches | % Correct Statement Matches |
|---|---|---|
| **WMD-GloVe** | **59** | **100%** |
| WMD-FastText | 58 | 98.31% |
| WMD-Word2Vec | 56 | 94.92% |
| PV + DBOW | 47 | 79.66% |
| PV-DM | 25 | 42.37% |
| PV-DBOW | 23 | 38.98% |
| LSA | 5 | 8.47% |

However, if we analyse performance in terms of the number of top three ranked returned statement hits, a subtle difference in the accuracy performance of WMD-GloVe and WMD-FastText compared to WMD-Word2Vec emerges. The WMD-GloVe system outperforms all other systems in the top 3 ranking comparisons by returning an impressive 100% (59/59) top 3 ranked search results for all possible search term combinations compared to

58/59 (98.31%) for the WMD-FastText system and 56/59 (94.92%) for the WMD-Word2Vec system. The robust ranking performance of all three WMD systems is contrasted against a significant drop in performance of the Doc2Vec models when the percentage of top 3 ranking returned statements is considered. The poor performance of the Doc2Vec architectures on shorter query-statement combinations ties in with evidence from other research that the paragraph embedding approaches are best suited to longer text segments (Galke et al. 2017). Le and Mikolov (2014), the developers of Doc2Vec, note that very short text tends not to generate useful representations from this model. If performance on shorter paragraphs or sentences is important, they suggest factoring in some mechanism to overweight them. The authors propose a method of repeating a paragraph that is 1/Nth the average size by N times randomly throughout the training set or implementing N times more steps during inferencing. The mediocre performance of the LSA system is further compounded with the lowest return of 5 statement matches (8.47%) in the top 3 ranked returned results. All WMD systems recorded strong performances with the top 30 and top three ranked comparisons.

*Table 3: The number and percentage of correct statement matches returned within the top ranked - #1 similarity ranking positions by each semantic retrieval model across 12 query-statement trials totalling 59 comparison statements.*

| System | # Correct Statement Matches | % Correct Statement Matches |
|---|---|---|
| **WMD-GloVe** | **53** | **89.83%** |
| **WMD-FastText** | 50 | 84.73% |
| **PV + DBOW** | 39 | 66.10% |
| **WMD-Word2Vec** | 18 | 58.98% |
| **PV-DBOW** | 16 | 27.12% |
| **PV-DM** | 15 | 25.42% |
| **LSA** | 2 | 0.03% |

However, if we analyse performance in terms of the number of top ranked (number one) returned statement hits, we begin to see a significant difference in the accuracy performance of WMD-GloVe and WMD-FastText compared to WMD-Word2Vec. The WMD-GloVe system returns an impressive 53/59 (89.83%) top ranked statement matches compared to a 50/59 (84.74%) return rate for the WMD-FastText system and a significantly lower 18/59 (58.98%) return rate for the WMD-Word2Vec system. In fact, the WMD-Word2Vec system is outperformed in these trials by the Doc2Vec PV+DBOW system which returns 39/59 (66.10%) top ranked statement matches. Through more detailed analysis, the superior performance of the WMD-GloVe system is further highlighted if we look at the results for the percentage of top 2 ranked results returned by each system. The WMD-GloVe system returns all search matches (100%) in the top 2 ranked positions compared to 56/59 (94.92%) top 2 ranked returns for WMD-FastText, 51/59 (86.44%) top 2 ranked returns for WMD-Google and 44/59 (74.58%) top 2 ranked returns for the Doc2Vec PV+DBOW system.

From the ranked similarity results it has been established that WMD-GloVe achieves the most robust accuracy performance for statement matching from re-worded and partial queries. The real power of this vector-based approach however, is its ability to achieve semantic matching of statements that are closely related in theme or content. To highlight this semantic processing ability, an informal qualitative analysis of the results of trial 2 was carried out to assess the relatedness of the top 20 statements that were returned based on the following query: "Michael Brown is the Chief Executive Officer of the Company". This query search was analysed as the WMD-based systems were the only models that achieved 100% recall on it. Using the search phrase, 10 out of the top 20 ranked statements (including the top 4 ranked statements) specifically mention "Michael Brown" as the CEO or executive director of the company. While this indicates successful term matching at a syntactic level, the contents of the remaining 10 statements demonstrate text matching at a semantic level by the combined WMD and GloVe word embeddings model. The remaining 10 statements reference semantically similar topics or concepts including management structure, key management personnel, board of directors and other details relating to company management. This ability to cluster semantically related concepts within the distribution space has important implications for facilitating users' search experiences through informed query expansion and relevant responses to partial or incomplete query searches.

In these experimental trials, pre-trained GloVe vectors (from Common Crawl Dataset) were used. They proved successful in accurate retrieval of both short and long query-statement searches. It should also be noted that, if domain specific tuning is necessary for specialised datasets, the GloVe model is much more efficient to train compared to the Doc2Vec variations. Scalability is facilitated as it is easier to parallelise the implementation enabling it to train over more data and populating the co-occurrence matrix requires a single pass through an entire corpus to collect the statistics.

*5.1 Summary*

WMD is lacking in full-text retrieval when relatively short queries were used to retrieve full documents. This was due to the mismatch in semantic space between the queries that consist of a small number of words within a restricted context compared a longer document that is rich with words in context. However, given the nature of this statement retrieval exercise, the queries and target statements share similar word lengths and will be closer to each other in semantic space (Kim et al. 2017). WMD proved to be very effective and accurate on the small dataset that was used in this experiment. However, scalability to a larger dataset may prove more challenging given that original WMD algorithm has a high time complexity of $O(n^3 \log n)$ where n represents the number of unique words in documents (Pele & Werman 2009). While a number of researchers have ruled out the use of the original, full implementation of WMD on full-text retrieval of large datasets (Galke, L; Saleh, A; Scherp 2017), a number of studies have reported promising accuracy levels with the implementation of a more computationally efficient 'Relaxed' Word Mover's Distance (RWMD) algorithm (Pele & Werman 2009; Kusner et al. 2015; Kim et al. 2017). Using a stripped back, efficient version of this algorithm, this similarity mechanism can be easily added as a feature to any relevance search framework to improve ranking performance without adding cumbersome overhead to the system. Another interesting development in the enhancement of efficiency and scalability of semantic similarity searching, are solutions that deploy word embeddings over traditional inverted-index search engines, thus taking advantage of this long established and robust technology. With these dense vectors being indexed and queried in text domains, authors report major efficiency gains and significantly faster search times with limited impact to retrieval accuracy (Růžička et al. 2017).

Our results show that pre-trained word embedding models can easily be applied to different domains with considerable success. However, there are certain use cases and business domains where the language used is particularly nuanced and specific. In these circumstances being uncoupled from the domain ontologies would be a disadvantage leading to systems that fail to understand the information needs of the end-users. It is therefore necessary to adopt embedding schemes that are enriched by exploiting domain tailored knowledge. The benefit of the unsupervised learning algorithms evaluated in this paper is that they can be trained on domain-specific ontology and lexical resources without the need of time consuming supervised learning prerequisites such as term extractors or manually labelled training data. The power of these algorithms is that their term x object matrices can be automatically populated for any text collection where the underlying concepts can be identified by completely automatic statistical processes.

## 5. Conclusion

The accuracy and robustness of these vector-based semantic retrieval models has set the agenda for this analysis paper. However, these mechanisms also need to be considered within a wider context of a semantic search framework and a semantic support infrastructure where usability and the support of the end-user are the focus of industrial information retrieval and management solutions. Given that a substantial performance gap remains between Information Retrieval systems and what users need and expect from them (van Opijnen & Santos 2017) and considering the volume of time that users can spend on searching tasks as part of their everyday work tasks, it is necessary to consider an entire search eco-system built around semantic search. Many large businesses have complex information spaces that require additional support for the end user to facilitate their search strategies for navigating these unfamiliar underlying knowledge structures. This is particularly necessary if they are seeking information from a wide range of topics or have uncertainty about the nature of the search problem. This lack of definition and certainty about the keywords to choose further highlights the need to factor semantic synonymy into the equation. Semantic search will be at the core of these new breed of techniques that are being developed to support the browsing of information spaces. Apart from text-based semantic search, which has been at the centre of this investigation, Information Retrieval accuracy can be further improved when we incorporate the semantic based approaches into a hybrid framework of search strategies.

Other measures of algorithmic relevance can be included into the overall suite of search tools including network statistics, click-through data and semantic driven query expansion (van Opijnen & Santos 2017). Query expansion is an iterative and exploratory process where the user actively engages with the search system to refine their queries in response to the results that are returned. This can be seen on the interfaces of many web search engines where expanded query suggestions appear in response to users' input. Use of query expansion has been found to increase recall and this process could be enriched through semantic based query recommendations or auto-suggest. Here the query expansion system would use the word embedding techniques to suggest potential query terms based on semantic similarity or synonymous names for concepts (Uren et al. 2007; Kuzi et al. 2016).

This investigation demonstrates the ability of semantic or conceptual-based search strategies to exploit the latent underlying semantic structure of text and how this can be leveraged to improve the quality and relevancy of the search experience. Semantic search enables the retrieval of documents by how similar the concepts in the query are to the concepts in the document. These concepts represent high-level ideas in a given domain. Semantic representation strategies can be viewed as a means of narrowing the gap between the mismatch of words that are contained in documents and the words expressed in queries as users' intentions. For the modern user, this intuitive behaviour of semantic search has almost become expected, thanks to services such as Google, as searchers expect search engines to 'do as I mean – not as I say' when they query. Semantic search enables users to get relevant results even when they input short-hand, truncated or misspelled queries containing only a few keywords. Overcoming lexical problems such as misspellings and partial queries facilitates data exploration and enables users to find target text in large collections of data allowing them to interact more fully with the data and enriching the information discovery process.

***This research was supported and co-authored by Allen and Overy Ltd as part of the work of the Ulster University Legal Futures Research. This centre provides research, development and educational resources for the promotion of innovation in legal services provision.***